\documentclass{amsart}

\newtheorem{theorem}{Theorem}[section]

\theoremstyle{definition}

\newtheorem*{7.1}{Theorem 7.1}
\newtheorem*{9.1}{Theorem 9.1}
\newtheorem*{9.2}{Theorem 9.2}
\theoremstyle{remark}
\newtheorem{remark}[theorem]{Remark}

\numberwithin{equation}{section}

\begin{document}
\title{Simplifying Local BRST Cohomology Calculation via Spectral Sequence}
\author{An Huang}
\address{Department of Mathematics, UC Berkeley}
\email{anhuang@berkeley.edu}
\date{August 08, 2010}

\begin{abstract}
\noindent We simplify some crucial calculations in \cite{1} by using the technique of spectral sequence of a double complex.
\end{abstract}
\maketitle
\section*{Introduction}
The calculation of local BRST cohomologies is important in many aspects in gauge theories. \cite{1} provides a nice review of results and their proofs on this topic. We found that some of the crucial mathematical calculations can be done much easier by using the technique of spectral sequence of a double complex. The main point of this paper is to make \cite{1} easier to access by mathematicians. With this goal in mind, in section 1, we will give a very short review of some background of \cite{1}. This review is neither intended to be comprehensive nor precise in details, however, we hope to explain the main mathematical points of \cite{1} and our paper as clearly and quickly as possible. In section 2, we copy and prove theorems 7.1, 9.1, and 9.2 of \cite{1} by using spectral sequence. Necessary background on spectral sequence can be found in \cite{2}, for example.
\section{Background And Notation}
In the BV formalism, the BRST symmetry generator is a differential $s$ called BRST differential acting on local differential forms on the jet space $\Omega$ of all relevant fields (fields, ghosts, and antifields) on spacetime of dimension $n$. The complex vector space $\Omega$ has a bigrading given by the pure ghost number $puregh$, and the antifield number $antifd$, which combine to give a total grading as the ghost number defined as $gh=puregh-antifd$. For convenience of the terminology of spectral sequence, in this paper, we use $-antifd$ instead of $antifd$ to denote the grading according to the antifield number. $s$ has a decomposition as the sum of two anticommuting differentials on $\Omega$ as $s=\delta+\gamma$, where $\delta$ increases $-antifd$ by $1$ while leaving $puregh$ unchanged, and $\gamma$ increases $puregh$ by $1$ and leaves $-antifd$ unchanged. Thus $s$ increases the $gh$ by $1$. Furthermore, all these differentials anticommute with the exterior derivative $d$. $\delta$ equals to zero when restricted to elements in $\Omega$ of antifield number zero, while $\gamma$ acts on these elements as a "gauge transformation where the gauge parameters are replaced by the ghosts". The cohomologies of $d$ and $\delta$ are not hard to calculate, and the results are summarized in chapters 4 and 5 of \cite{1}, respectively. The cohomology of $r$ can be calculated by Lie algebra cohomology, and relevant results are summarized in chapter 8. The theoretical calculation of the cohomologies of $s$ or $s|d$ (the so called BRST cohomologies) based on these results are discussed in chapters 7 and 9. This is the part where we find the most important results can be obtained much easier by the technique of a spectral sequence of a double complex. Hopefully our treatment also makes it clearer when results of these theorems hold in certain different cases. We will give some remarks where such considerations are actually useful.
\bigskip\\
{\bf Notation.}\\
$H(w|d)$: If $w$ is a differential on $\Omega$ anticommuting with $d$. Then $\Omega/d\Omega$ is also a complex with $w$ acting as the differential. $H(w|d)$: denotes the cohomology of this complex when $\Omega$ is clear from the context.\\
$H^g_0(\gamma,H(\delta))$ (or $H^{g,p}_0(\gamma,H(\delta|d))$): They represent cohomologies of the complex $H(\delta)$ (or $H(\delta|d)$) with differential $\gamma$ at degrees indicated as in theorem 7.1 of \cite{1}.
\section{Spectral Sequence Proofs of Theorems 7.1, 9.1, and 9.2 in \cite{1}}
\begin{7.1}
In the space of local forms, one has the following isomorphisms:\\
\begin{equation}\label{A}
H^g(s)\cong H^g_0(\gamma,H(\delta))
\end{equation}
\begin{equation}\label{B}
H^{g,p}(s|d)\cong
\begin{cases}
H^{g,p}_0(\gamma,H(\delta|d)) & \text{if } g\geq 0\\
H^p_{-g}(\delta|d) & \text{if } g < 0.
\end{cases}
\end{equation}
where the superscripts $g$ and $p$ indicate the (total) ghost number and the form-degree respectively and the subscript indicates the antifield number.
\end{7.1}
\begin{proof}
$\Omega$ becomes a double complex with the bigrading given by $puregh$, $-antifd$, together with the horizontal differential $\gamma$ and the vertical differential $\delta$. By discussions in section 7.3 of \cite{1}, we may assume that for any cohomology class in $H^g(s)$ with a given ghost number $g$, $-antifd$ is bounded from below. Therefore if one can show that the cohomology of the double complex at ghost number $g$ is stable with respect to the choice of a lower bound for $-antifd$, it must be equal to $H^g(s)$. Choose any lower bound for $-antifd$ less than zero, by theorem 5.1, the only possibly nonzero terms in $E_2$ of the spectral sequence are the column with $-antifd=0$, and these terms are $H^{g,0}_{\gamma}H_{\delta}(\Omega)$. Therefore $\Omega$ degenerates at $E_2$. For each ghost number $g$, $H^{g,0}_{\gamma}H_{\delta}(\Omega)$ is the only possibly nonzero term in $E_{\infty}$. Note that $\delta$ restricted to $-antifd=0$ piece of $\Omega$ is zero, hence $H^{g,0}_{\gamma}H_{\delta}(\Omega)$ is the same as $H^g_0(\gamma,H(\delta))$. Therefore the cohomology of the double complex is independent of the choice of the lower bound for $-antifd$, it equals $H^g(s)$, and $H^g(s)\cong H^g_0(\gamma,H(\delta))$.\\
For \eqref{B}, since the differentials $s$, $\delta$ and $\gamma$ all anticommute with $d$, the quotient $\Omega/d\Omega_F$ is also a double complex with the same bigrading and differentials $s=\delta+\gamma$. For any given ghost number $g$, for elements in $H^{g,p}(s|d)$, the same consideration for lower bounds of $-antifd$ goes through. Choose any lower bound for $-antifd$ less than $\text{min}\left\{0,g\right\}$, we note that from corollary 6.1 of \cite{1}, $H^p_k(\delta|d,\Omega^*_{N_C>0})$ vanish for all $k\geq1$. Therefore, possibly nonzero terms in $E_2$ of this spectral sequence either has pure ghost number equal to $0$, or has antifield number equal to $0$. If $g\geq 0$, again keeping in mind that $\delta$ restricted to $-antifd=0$ piece of $\Omega$ is zero, the only term with this property is equal to $H^{g,p}_0(\gamma,H(\delta|d))$. Otherwise, it is $H^{g,p}_{-g,\gamma}H_{\delta|d}(\Omega)$ ($H_{\gamma}H_{\delta|d}(\Omega)$ at $gh=g$, $-anti=g$, form degree=$p$). Since $\gamma$ increases the pure ghost number by $1$, $\gamma$ maps $H_{\delta|d}(\Omega)$ to a cohomology group of $\delta|d$ with positive pure ghost number, thus zero. Hence $H^{g,p}_{-g,\gamma}H_{\delta|d}(\Omega)=H^p_{-g}(\delta|d)$. So \eqref{B} holds.
\end{proof}
\begin{remark}
From the above proof it is clear that \eqref{A} can hold in a double complex other than $\Omega$, provided that one has an appropriate generalization of theorem 5.1. Understanding this point is relevant to chapter 8 of \cite{1}.
\end{remark} 
\begin{9.1} 
If $H^p(d,\Omega)=\delta^p_0\mathbb{R}$ for $p=0,...,n-1$ and the equations of motions are consistent, there exists a basis
\begin{equation}
\left\{[1],[h^0_{i_r}],[\widehat{h}_{i_r}],[e^0_{\alpha_s}]\right\}
\end{equation}
of $H(s)$ such that the representations fulfill $sh^{r+1}_{i_r}+dh^r_{i_r}=\widehat{h}_{i_r}, sh^{r}_{i_r}+dh^{r-1}_{i_r}=0,..., sh^{1}_{i_r}+dh^0_{i_r}=o, sh^0_{i_r}=0$\\
and\\ $\text{form degree } e^s_{\alpha_s}=n, se^s_{\alpha_s}+de^{s-1}_{\alpha_s}=0,..., se^1_{\alpha_s}+de^{0}_{\alpha_s}=0, se^0_{\alpha_s}=0$\\ 
for some forms $h^q_{i_r}$, $q=1,...,r+1$ and $e^p_{\alpha_s}$, $p=1,...,s$. Here, $[a]$ denotes the class of the s-cocycle $a$ in $H(s)$.
\end{9.1}
\begin{proof}
We endow the vector space $\Omega$ with a new bigrading given by the ghost number and the form degree. Then $\Omega$ becomes a double complex with horizontal differential $d$ and vertical differential $s$. As vector spaces, we have 
\begin{equation}\label{R}
E_i\cong \text{Ker}(d_i)\oplus E_i/\text{Ker}(d_i)\cong E_{i+1}\oplus d_iE_i\oplus E_i/\text{Ker}(d_i)
\end{equation}
as by definition, $E_{i+1}=H(d_i)$.\\ 
In addition we have $E_1=H(s)$. Since the form degree is bounded by $n$, for each total degree the ghost number is bounded from below, and the spectral sequence degenerates at $E_{n+1}$. Apply \eqref{R} repeatedly for $i=1,2,...,n$, we see that $H(s)$ is isomorphic to the direct sum of $E_{n+1}$ together with $d_iE_i\oplus E_i/\text{Ker}(d_i)$ for $i=1,2,...,n$. $1$ and the $e^0_{\alpha_s}$ are elements of $E_{n+1}$, since they can be extended to a zig-zag of maximal length. $1$ is singled out, and we complete a basis for $E_{n+1}$ and denote other basis elements as $e^0_{\alpha_s}$. The $\widehat{h}_{i_r}$ and $h^0_{i_r}$ are elements of a basis of $d_iE_i$ and $E_i$, respectively.
\end{proof}
\begin{remark}
From the proof we see that the conditions that $H^p(d,\Omega)=\delta^p_0\mathbb{R}$ for $p=0,...,n-1$ and the equations of motions are consistent are not really needed for the above theorem to be true. However, they become important for the next theorem to be true.
\end{remark} 
\begin{9.2}
If $\left\{[1],[h^0_{i_r}],[\widehat{h}_{i_r}],[e^0_{\alpha_s}]\right\}$ is a basis of $H(s)$ with the properties of theorem 9.1, then an associated basis of $H(s|d)$ is given by 
\begin{equation}
\left\{[1],[h^q_{i_r}],[e^p_{\alpha_s}]: q=0,...,r, p=0,...,s\right\} 
\end{equation}
where in this last list, $[]$ denotes the class in $H(s|d)$.
\end{9.2}
\begin{proof}
Let $a$ be an element in $H(s|d)$. According to discussions in section 9.2 of \cite{1}, when $H^p(d,\Omega)=\delta^p_0\mathbb{R}$ for $p=0,...,n-1$ and the equations of motion are consistent, we have the descent equations for $a=\omega^l$: $s\omega^l+d\omega^{l-1}=0$, $s\omega^{l-1}+d\omega^{l-2}=0$,..., $s\omega^0=0$. Therefore $\omega^0$ is an element in $H(s)$, and as an element in $H(s)$ it equals a linear combination of $[1],[h^0_{i_r}],[\widehat{h}_{i_r}],[e^0_{\alpha_s}]$. Note $a=\omega^l$, as an element in $\Omega$, is determined by $\omega^0$ up to an element in $H(s)$. Therefore $a$ as an element in $\Omega$ equals a linear combination of $1,h^q_{i_r},e^p_{\alpha_s}: q=0,...,r, p=0,...,s$ plus a linear combination of $1,h^0_{i_r},\widehat{h}_{i_r},e^0_{\alpha_s}$ plus an $s$ exact term. Since the elements $\widehat{h}_{i_r}$ are trivial in $H(s|d)$, we conclude that the cohomology class $a$ in $H(s|d)$ equals a linear combination of elements $[1],[h^q_{i_r}],[e^p_{\alpha_s}]: q=0,...,r, p=0,...,s$.\\
As for  showing that $[1],[h^q_{i_r}],[e^p_{\alpha_s}]: q=0,...,r, p=0,...,s$ are linearly independent in $H(s|d)$, we have little to add and refer the reader to the proof of this point in \cite{1}, pages 79-80.
\end{proof}  
\begin{remark}
The equations of motion are consistent has an important implication for the descent equations, which is used implicitly in the above proof. For the explanation of this point, the reader can consult page 71 of \cite{1}.
\end{remark}
\section*{Acknowledgments}
I thank my advisor Richard Borcherds for suggesting me this topic.
\bibliographystyle{amsplain}

\end{document}